\documentclass[preprint,floats,aps,epsfig,nofootinbib,amssymb]{revtex4-1}
\usepackage{mathrsfs}

\usepackage{slashed}
\usepackage{graphicx,color}
\usepackage{dcolumn}
\usepackage{bm}
\usepackage{graphicx}
\usepackage{amssymb}

\begin{document}

\title{Vacuum Stability and Higgs Diphoton Decay Rate in the Zee-Babu Model }

\author{Wei Chao${^{1,2}}$}
\email{chaow@physics.wisc.edu}

\author{Jian-Hui Zhang$^1$}
\email{zhangjianhui@gmail.com}
\author{Yongchao Zhang $^3$}
\email{yczhang@pku.edu.cn}

 \affiliation{$^1$INPAC, Shanghai Jiaotong University, Shanghai, 200240, China\\
  $^2$Department of Physics, University of Wisconsin-Madison, Madison, WI 53706, USA\\  $^3$ Center of High Energy Physics, Peking University.
 }

\begin{abstract}
Although recent Higgs data from ATLAS and CMS are compatible with a Standard Model (SM) signal at $2\sigma$ level, both experiments see indications for an excess in the diphoton decay channel, which points to new physics beyond the SM. Given such a low Higgs mass $m_H \sim 125~{\rm GeV}$, another sign indicating the existence of new physics beyond the SM is the vacuum stability problem, i.e., the SM Higgs quartic coupling may run to negative values at a scale below the Planck scale. In this paper, we study the vacuum stability and enhanced Higgs diphoton decay rate in the Zee-Babu model, which was used to generate tiny Majorana neutrino masses at two-loop level. We find that it is rather difficult to find overlapping regions allowed by the vacuum stability and diphoton enhancement constraints. As a consequence, it is almost inevitable to introduce new ingredients into the model, in order to resolve these two issues simultaneously.  
\end{abstract}

\draft

\maketitle
\section{Introduction}

In the Standard Model~(SM) of particle physics, the Higgs mechanism provides an explanation to the spontaneous electroweak symmetry breaking, but the Higgs boson itself left no trace in all the previous high-energy collider  experiments. Recently both ATLAS and CMS collaborations announced an observation of a Higgs-like boson at the $5\sigma$ confidence level~\cite{Giannotti}. The discovery points to a new resonance near the $125~{\rm GeV}$, which,  if confirmed to be the SM Higgs boson, will be a milestone for fundamental particle physics. We assume it is the SM Higgs boson in this paper. Such a low Higgs mass then immediately leads to problems for vacuum stability, which requires the Higgs self-coupling $\lambda$ remain positive at all scales.  Given $M_H\sim 125~{\rm GeV}$, the Higgs self-coupling $\lambda$ would turn negative at a scale $\Lambda\sim 10^9 -10^{11}~{\rm GeV}$~\cite{instability}, indicating the existence of new physics beyond the SM around that scale.

It's quite interesting to notice that both the ATLAS and CMS collaborations report a significant enhancement in the diphoton decay channel
\begin{eqnarray}
R_{\gamma\gamma} =
{ [\sigma(gg\to h) \times {\rm BR } (h\to \gamma\gamma) ]_{\rm obs} \over  [\sigma(gg\to h) \times {\rm BR } (h\to \gamma\gamma) ]_{\rm SM}  }
=1.71\pm0.33 \; ,
\end{eqnarray}
while $\sigma(gg\to h ) \times {\rm BR } (h\to ZZ^*)$ and $\sigma(gg\to h ) \times {\rm BR } (h\to WW^*)$ seem compatible with the SM predictions~\cite{Giannotti,Baglio:2012et}. In the absence of direct signals of new physics at colliders, the enhancement of $ \Gamma (h\to \gamma\gamma)$ is an important hint of underlying new physics, since the Higgs boson couples to photons indirectly via loops where new physics can enter.

In general, the running of the Higgs self-coupling may be changed by the following types of interactions: quartic scalar interaction, Yukawa interaction and gauge interaction. One possibility to resolve the SM Higgs vacuum stability problem is to extend the SM with additional quartic scalar interactions associated with new scalar degrees of freedom. Typical examples are the Higgs-portal dark matter models, e.g. the darkon~\cite{darkon} and inert~\cite{inert} dark matter models. For a detailed analysis of the implications of these models on Higgs vacuum stability, see Refs.~\cite{raidal,taiwan,higgsportal,threshold,gondering,Chun:2012jw, Cheung:2012nb,Chao:2012mx}. Another possibility to improve the SM Higgs vacuum stability is to introduce new gauge degrees of freedom, in this case one may require new electroweak multiplets, the neutral component of which can be cold dark matter candidate, and (or) a new $U(1)$ gauge symmetry~\cite{Chao:2012mx}, in which the SM Higgs carries non-zero charge. To explain the enhanced $H\to \gamma \gamma $ decay width one needs to extend the SM with charged particles~\cite{Carena:2012xa}, which can be charged scalars in the type-II seesaw models~\cite{seesawII}, or multi-charged components in a septuplet dark matter model~\cite{Cai:2012kt} that couple to the SM through the quartic scalar interaction, or charged vector-like fermion doublets that have Yukawa interactions with the SM Higgs, or new $W^\prime$ vector boson which may couple to the SM Higgs through the kinematic term or through the  $W^{\prime\dagger} W^\prime H^\dagger H $ term. There have been extensive studies focusing on this topic~\cite{hrr,hrrHTM,hrrHTM2,hrrHTMWang,Chang:2012ta,Dorsner:2012pp}.

On the other hand, there are other definitive evidence of new physics beyond the SM, including neutrino masses and dark matter. The solar, atmospheric, reactor and accelerator neutrino experiments have provided convincing evidence that  neutrinos are massive and lepton flavors are mixed. In order to generate tiny neutrino masses, one may extend the SM with right-handed neutrinos that have large Majorana masses~\cite{seesawI}. Through Yukawa interactions the right-handed neutrinos couple to the SM lepton doublets, three active neutrinos then acquire tiny Majorana masses as given by the Type-I seesaw formula: $M_\nu=-M_D M_R^{-1} M_D^T $, where $M_\nu $ is the mass matrix of light active neutrinos, $M_D$ is the neutrino Dirac mass matrix linking the left-handed active neutrinos with the right-handed neutrinos, $M_R^{}$ is the mass matrix of right-handed neutrinos. Actually there are three types of tree-level seesaw scenarios~\cite{seesawI,seesawII,seesawIII} as well as three radiative seesaw scenarios~\cite{Ma:1998dn,zee,zee-babu}, which may generate Majorana masses for active neutrinos, with the help of dimension-five Weinberg operator~\cite{Weinberg-operator}.

In this paper, we study the Higgs vacuum stability problem and the enhanced $H\to \gamma \gamma $ decay rate in a radiative neutrino mass model, i.e. the minimal Zee-Babu model~\cite{zee-babu}, in which two $SU(2)_L^{}$ singlet scalar fields $\phi^+$ and $\kappa^{++}$ are introduced with hypercharge $1$ and $2$, respectively. The scalar potential is given as follows~\cite{zee-babu}
\begin{eqnarray}
V_H &= & (V_H)_{\rm SM}^{} +M_2^2 \phi^\dagger \phi +M_3^2 \kappa^\dagger \kappa + \lambda_1 (\phi^\dagger \phi)^2 + \lambda_{2} (\kappa^\dagger \kappa)^2 + \lambda_3^{} (\phi^\dagger \phi ) (H^\dagger H) \nonumber \\&&+\lambda_4^{} (\kappa^\dagger \kappa ) (H^\dagger H) + \lambda_5^{} (\phi^\dagger \phi ) (\kappa^\dagger \kappa) + (\mu \phi^-\phi^- \kappa^{+ } + {\rm h. c.} ) \; , \label{lagrangianzb}
\end{eqnarray}
where $(V_H)_{\rm SM}$ is the SM Higgs potential. The new Yukawa couplings linking the charged scalars to the leptons are
\begin{eqnarray}
{\cal L}_Y &=& f_{\alpha \beta } \overline{\ell_{L\alpha}^C} \varepsilon \ell_{L\beta}^{} \phi^+ + g_{\alpha \beta } \overline{E_{R\alpha}^C } E_{R \beta}^{} \kappa^{++} + h.c. \; ,
\end{eqnarray}
where $\ell_L$ denote left-handed lepton doublets, $E_R^{}$ are right-handed charged leptons.  The trilinear $\mu$ term in Eq. (\ref{lagrangianzb}) breaks lepton number explicitly, one naturally expects it to be reasonably small, since the symmetry is enhanced in the limit $\mu\rightarrow 0$.  Light neutrino masses are generated at two-loop level, which is
\begin{eqnarray}
 m_\nu = 16 \mu f_{ac}^{} m_c g^*_{cd} I_{cd} m_d f_{bd}^{}\ ,
\end{eqnarray}
where $m_c$ are charged lepton masses and $I_{ij}$ is a two-loop integral~\cite{zee-babu-loop}. It turns out that this model is not only capable of generating neutrino mass, the interaction terms proportional to $\lambda_3$ and $\lambda_4$ also provide a possibility of resolving the SM Higgs vacuum stability problem and explaining the enhanced $H\to \gamma \gamma $ decay rate. 

The paper is organized as follows: In section~\uppercase\expandafter{\romannumeral2}, we study numerically the SM Higgs vacuum stability and the Zee-Babu scalar potential vacuum stability. Section~\uppercase\expandafter{\romannumeral3} is devoted to investigating the enhanced $H\to \gamma \gamma$ decay rate induced by the charged scalar singlets. We investigate briefly the searches of the new charged scalars at the LHC in section~\uppercase\expandafter{\romannumeral4}, and summarize in section~\uppercase\expandafter{\romannumeral5}.

\section{Vacuum stability}

A constraint on the Higgs mass can be obtained by the requirement that spontaneous symmetry breaking actually occurs, that is, $V(v)$ is the minimum of the Higgs potential, where $v\approx 246~{\rm GeV}$ is the Higgs vacuum expectation value (VEV) .  The one-loop  Renormalization Group Equation (RGE) for the quartic coupling, including fermion and gauge boson contributions, is~\cite{RGE}
\begin{eqnarray}
16 \pi^2 \beta_\lambda^{(1)} =+12 \lambda^2 -\left(  {9 \over 5} g_1^2 + 9 g_2^2 \right) \lambda +
{9 \over 4 } \left(  { 3 \over 25 } g_1^4 + {2 \over 5 } g_1^2 g_2^2 +g_2^4 \right) + 12 \lambda y_t^2  -12 y_t^4 \; ,
\end{eqnarray}
where the top quark Yukawa coupling is given by $ y_t =\sqrt{2} M_t/v$ with $M_t$ the running mass of the top quark. If the initial value of the coupling $\lambda$ is too small, the top quark contribution can be dominant and drive it to a negative value $\lambda(\Lambda) < 0$, leading to the scalar potential $V(\Lambda )<V(v )$. The vacuum is not stable anymore since it has no minimum. Given $M_H\sim 125 $~${\rm GeV}$, the Higgs self-coupling $\lambda$ would run negative at a scale $\Lambda \sim 10^9 - 10^{11}~{\rm GeV}$~\cite{instability}.

We now analyze the vacuum stability and perturbativity constraints in the Zee-Babu model. There are several new parameters, $\lambda_1$, $\lambda_2$, $\lambda_3$, $\lambda_4$, $\lambda_5$, $g$ and $f$, which potentially affect the RGE running of the Higgs quartic coupling.  It turns out that both $\lambda_3$ and $\lambda_4$ play an important role through their contributions to the RGE of the Higgs quartic coupling. These couplings can alleviate both the vacuum stability and perturbativity constraints on the SM Higgs mass. In our analysis we employ the two-loop RGEs for the SM couplings~\cite{RGE} and one-loop RGEs for the new couplings associated with the Zee-Babu model. The one-loop $\beta$-functions of the couplings in the scalar potential can be written as
\begin{eqnarray}\label{betafuns}
16\pi^2 \beta_{\lambda_{~}} &=& {9 \over 4 } \left(  { 3 \over 25 } g_1^4 + {2 \over 5 } g_1^2 g_2^2 +g_2^4 \right) -\left(  {9 \over 5 } g_1^2 + 9 g_2^2 \right) \lambda +12 \lambda^2 +2\lambda_3^2 + 2 \lambda_4^2 -12 y_t^4 +12 \lambda y_t^2 \; , \nonumber \\
16\pi^2 \beta_{\lambda_1 }^{} & =& {54 \over 25 } g_1^4 -{36 \over 5} g_1^2 \lambda_1 + 20 \lambda_1^2 +2 \lambda_3^2 + \lambda_5^2 - 2 {\rm Tr} [(f^\dagger f)^2 ] +   2 {\rm Tr} [f^\dagger f]  \lambda_1  \; , \nonumber\\
16\pi^2 \beta_{\lambda_2 }^{} & =& {864 \over 25 } g_1^4 -{144 \over 5} g_1^2 \lambda_2 + 20 \lambda_2^2 +2 \lambda_4^2 + \lambda_5^2 - 2 {\rm Tr} [(g^\dagger g)^2 ] +   2 {\rm Tr} [g^\dagger g]  \lambda_1  \; , \nonumber\\
16\pi^2 \beta_{\lambda_3 }^{} & =& {27 \over 25 } g_1^4 -{9 \over 2} g_1^2 \lambda_3 - { 9 \over 2 } g_2^2 \lambda_3 + 6 \lambda \lambda_3 + 8 \lambda_1 \lambda_3  + 4 \lambda_3^2 +2 \lambda_4 \lambda_5 + 6 y_t^2 \lambda_3  +  {\rm Tr} [f^\dagger f]  \lambda_3  \; , \nonumber\\
16\pi^2 \beta_{\lambda_4 }^{} & =& {108 \over 25 } g_1^4 -{153 \over 10} g_1^2 \lambda_4 - { 9 \over 2 } g_2^2 \lambda_4 + 6 \lambda \lambda_4 + 8 \lambda_2 \lambda_4  + 4 \lambda_4^2 +2 \lambda_3 \lambda_5 + 6 y_t^2 \lambda_4  +  {\rm Tr} [g^\dagger g]  \lambda_4  \; , \nonumber\\
16\pi^2 \beta_{\lambda_5 }^{} & =& { 432 \over 25} g_1^4  - { 90 \over 5 } g_1^2 \lambda_5 + 4\lambda_3 \lambda_4 +8 \lambda_1 \lambda_5 + 8 \lambda_2 \lambda_5 +4 \lambda_5^2 + ( {\rm Tr} [f^\dagger f] + {\rm Tr} [g^\dagger g] )\lambda_5  \; .
\end{eqnarray}
We assume
$ {\rm Tr} [f^\dagger f],~ {\rm Tr} [g^\dagger g] \ll 1$, which is consistent with the requirements of the tiny Majorana neutrino masses,
such that we can safely neglect them in the numerical analysis.  $\phi$ and $\kappa$ contribute to the $\beta$-function of gauge coupling $g_1$, which can be written as
\begin{eqnarray}
\beta_{g_1}^{} = \left(  {41 \over 10 } +1 \right) g_1^3 \; ,
\end{eqnarray}
Here the SM gauge couplings are normalized based on $SU(5)$, so the electroweak couplings $g$, $g^\prime$ are related to $g_1, g_2$ via $g=g_2$ and $g^\prime = \sqrt{3/5} g_1$~\cite{3/5}. The couplings $\alpha_i\equiv g_i^2 /4 \pi$ are given as $(\alpha_1,~\alpha_2,~\alpha_3) = (0.01681,~0.03354,~0.1176)$ at the Z-pole~\cite{pdg2012}.  The full set of $\beta-$functions for gauge couplings, top quark Yukawa coupling; the matching condition between the top quark pole mass and running mass, and the one-loop matching condition  through which $M_H$ is determined by the relation to the running Higgs quartic coupling, can be found in Ref.~\cite{RGE}.  We set the running mass of the top quark to be $M_t =172.5~{\rm GeV}$ at the scale of $M_Z$ in our numerical analysis.

\begin{figure}[htbp]
\hspace*{1.5em}
\includegraphics[width=0.4\textwidth]{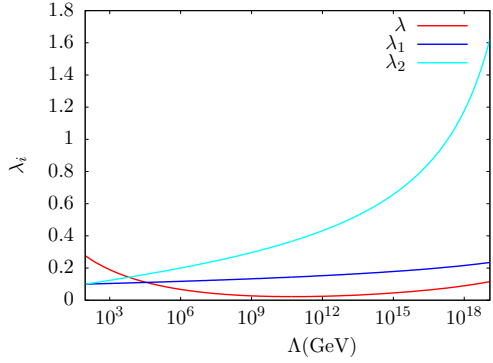}
\caption{Running of quartic scalar self-couplings in the Zee-Babu model as a function of the cut-off scale, where $\lambda_1=0.1, \lambda_2=0.1, \lambda_3=0.2, \lambda_4=0.4, \lambda_5=-0.2$ at the electroweak scale.}
\label{runninglambda}
\end{figure}

To ensure vacuum stability, the coefficients of quartic scalar couplings must be non-negative in any direction and at any renormalization scale. This leads to the following constraints on the scalar couplings:
\begin{itemize}
\item If  $\lambda_{3,4}>0$, the vacuum stability condition is
\begin{eqnarray}
\lambda, \lambda_{1,2}>0\; , \hspace{2em }  \lambda_5+\sqrt{4\lambda_1 \lambda_2}>0  \; .
\end{eqnarray}
\item If $\forall$  $\lambda_{3, 4} <0 $, the following additional constraints are needed
\begin{eqnarray}
&&\sqrt{2\lambda \lambda_1} - |\lambda_3|>0, \hspace{1em} \sqrt{2\lambda \lambda_2}- |\lambda_4|>0\  (\rm{for}\  A>0\  \rm{or}\  A<0\  \& \ {\frac{-B}{2A}}\notin[0,1]) ,  \nonumber \\
&&\sqrt{2\lambda \lambda_1} - |\lambda_3|>0, \hspace{1em} \sqrt{2\lambda \lambda_2}- |\lambda_4|>0, \hspace{1em} C-\frac{B^2}{4A}<0 \ (\rm{otherwise})\ ,
\end{eqnarray}
\end{itemize}
where $A=(\lambda_3-\lambda_4)^2-2\lambda(\lambda_1+\lambda_2-\lambda_5)$, $B=4\lambda\lambda_1-2\lambda_3^2+2\lambda_3\lambda_4-2\lambda\lambda_5$, $C=\lambda_3^2-2\lambda\lambda_1$. These constraints should be valid at any renormalization scale, and are obtained in analogy with the method used in Ref.~\cite{Kanemura:2000bq}. In our numerical analysis, we also assume the na\"ive perturbativity constraint, namely assume $\lambda < 8.2$~\cite{8.2} and $ |\lambda_i| < 4\pi$ $(i= 1\cdots 5) $ from the electroweak scale $\Lambda_{EW}$ to the cut-off scale $\Lambda_{c}$. In this analysis, we take the cut-off scale $\Lambda_{c}$ to be the Planck scale.

Note that the extra scalars $\phi, \kappa$ do not couple to quarks, therefore the running of their self-couplings do not receive negative contribution from the top Yukawa coupling. As a consequence, they can easily reach the Landau pole if their initial input values at the electroweak scale are too large. This is indeed what we found in our scan. For small values of $\lambda_{1,2}$, we found solutions rendering the vacuum stable at all scales up to the Planck scale. In Fig.~\ref{runninglambda} we plot as an illustrative example the running of the scalar quartic couplings in the Zee-Babu model. As the plot shows, after including the contribution of extra scalars $\phi, \kappa$ to the running, the Higgs quartic coupling $\lambda$, which turns negative at $10^9- 10^{11} {\rm GeV}$ in the SM, becomes positive up to the Planck scale. From our scan we find that for the vacuum to be stable, the couplings $\lambda_3$ and $\lambda_4$ tend to be of the same sign at the electroweak scale. As we will see in the next section when we discuss the enhancement of Higgs diphoton decay rate, this is not the phenomenologically favored case.

\section{Enhancement of the $h \rightarrow \gamma\gamma$ decay rate}

As we have mentioned in the introduction, the singly and doubly charged scalars $\phi$ and $\kappa$ in the Zee-Babu mdoel provide a possible explanation for the enhancement of the $h \rightarrow \gamma\gamma$ decay rate. More specifically, $\phi$ and $\kappa$ carry no color charge, implying their vanishing contribution to the gluon fusion production of Higgs, which is the dominate production mechanism at the LHC. However, due to the scalar interaction with Higgs, i.e. $-\lambda_3(\phi^\dagger\phi)(H^\dagger H) -\lambda_4(\kappa^\dagger\kappa)(H^\dagger H)$, these two new scalars can augment the Higgs diphoton decay rate to some extent. In this section we will show that the large enhancement ratio $R_{\gamma\gamma}$ recently observed at the LHC can be easily explained in the Zee-Babu model by the $\phi$ and $\kappa$ loops. To make our analysis solid and reliable, we will consider the constraints on the parameters from perturbativity
and collider search results.

\subsection{$\phi$ on, $\kappa$ off}

We first turn off the contribution from the scalar $\kappa$ artificially (or this can be understood as the limit $\lambda_4=0$ at the electroweak scale), and consider the effect of $\phi$ on the Higgs diphoton decay rate, to see if $\phi$ is sufficient to produce the large enhancement ratio
. The partial width for the diphoton channel in the SM is~\cite{htodiphoton}
\begin{eqnarray}
\Gamma(h\to \gamma\gamma)=\frac{G_{\rm F} \alpha^2 M_h^3}{128\sqrt{2}\pi^3}\left|A_1(\tau_W)+ N_C Q_t^2  A_{1/2}(\tau_t) \right |^2 \ ,
\end{eqnarray}
where $G_{\rm F}$ is the Fermi constant, $\alpha$ the fine-structure constant,
$N_C=3$ the number of colors, $Q_t=+2/3$ the top quark electric charge in units of $|e|$, $\tau_i\equiv 4M_i^2/M_h^2$, $i=t, W$, and the contribution from spin-0, $1/2$ and 1 particles are
\begin{eqnarray}
\label{As}
&& A_0(x) = -x^2 \left[\frac{1}{x}-f\left(\frac{1}{x}\right)\right]  \,, \\
&& A_{1/2}(x) = 2x^2 \left[\frac{1}{x} +\left(\frac{1}{x}-1\right)f\left(\frac{1}{x}\right)\right]  \,, \\
&& A_1(x) = -x^2 \left[\frac{2}{x^2} +\frac{3}{x} +3\left(\frac{2}{x}-1\right)f\left(\frac{1}{x}\right)\right]  \ ,
\end{eqnarray}
respectively. The function $f(x)$ is defined as
\begin{equation}
f(x)=\left\{\begin{array}{lcc}
[\,\arcsin\sqrt{x}\,]^2&\mbox{for}&x\leq 1\,,\\[5mm]
-{\displaystyle\frac{1}{4}\,\left[\ln\frac{1+\sqrt{1-x^{-1}}}{1-\sqrt{1-x^{-1}}}-i\pi\right]^2}
&\mbox{for}&x>1\,.\end{array}\right.
\end{equation}
For a 125~GeV Higgs, the contribution from the SM $W$ boson $A_1 = -8.32$ is the dominant part and interferes destructively with the contribution from the top quark $N_C Q_t^2A_{1/2} = 1.84$.

After electroweak symmetry breaking, the scalar $\phi$ acquires mass: $M_{\phi}^2 = M_2^2 + 1/2 \lambda_3 v^2 $, and its coupling to the SM Higgs is $g_{h\phi\phi^\dagger} = \lambda_3v$. With contribution from the new scalar $\phi$, the enhancement ratio of the Higgs diphoton decay rate is, compared to the SM value~\cite{Carena:2012xa}
\begin{eqnarray}
\label{phi}
R_{\gamma\gamma}^\phi = \left| 1+  \frac{1}2 \lambda_3 \frac{v^2}{M_\phi^2}\frac{A_0(\tau_{\phi})}{ A_1(\tau_W)+ N_c Q_t^2 \, A_{1/2}(\tau_t)}\right|^2\ .
\end{eqnarray}
It is obvious that the enhancement ratio is determined by two parameters in this case, the $\phi$-Higgs quartic coupling $\lambda_3$ and the $\phi$ mass $M_\phi$. We first consider the case $\lambda_3 < 0$, where the $\phi$-loop interferes constructively with the SM $W$ boson contribution and tends to enlarge the diphoton decay width.

\begin{figure}[t]
  \centering
  \includegraphics[width=8.3cm]{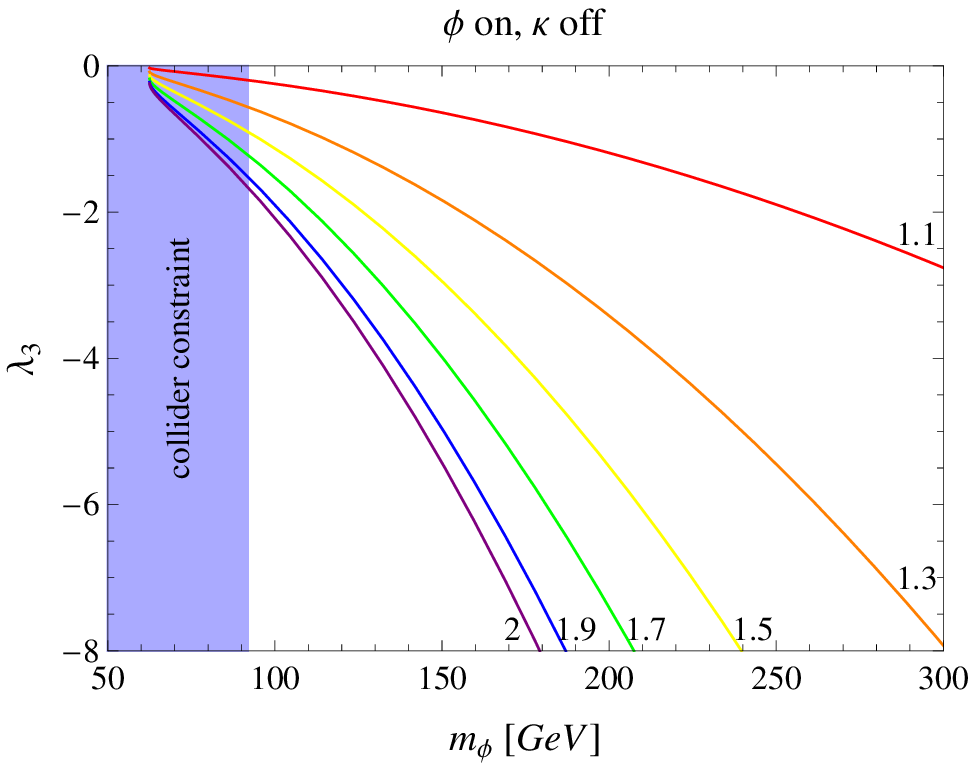} \hspace{-.5cm}
  \includegraphics[width=8.3cm]{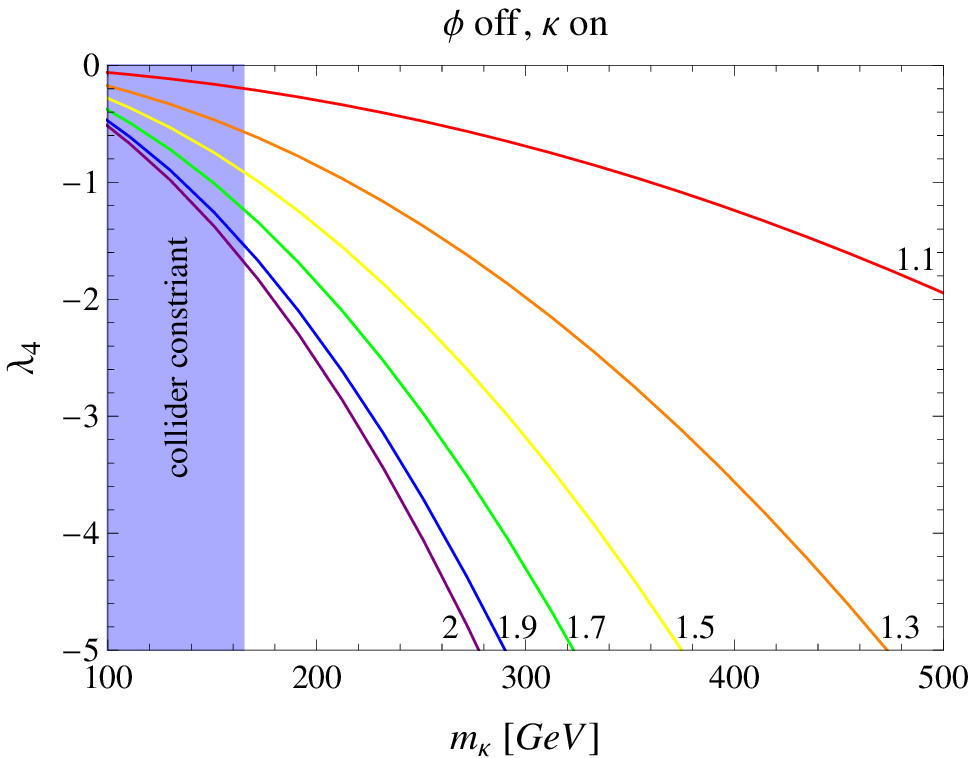}
  \caption{Left panel: Plots of the enhancement ratios
  $R_{\gamma\gamma}^{\phi} =1.1,\,1.3,\,1.5,\,1.7,\,1.9,\,2$ for $\Gamma(h \rightarrow \gamma\gamma)$ as functions of $\lambda_3$ and $M_\phi$, when we turn on only the scalar $\phi$ in the Zee-Babu model, in addition to the SM particles.
  Right panel: contours of $R_{\gamma\gamma}^{\kappa} =1.1,\,1.3,\,1.5,\,1.7,\,1.9,\,2$ as functions of $\lambda_4$ and $M_{\kappa}$, when we turn on only the scalar $\kappa$.
  The blue regions on the left side in the panels are the bounds of collider experiments on $M_\phi$ and $M_\kappa$, respectively. See text for details.}
  \vspace{-.3cm}
  \label{figphi}
\end{figure}

We should notice that both $\lambda_3$ and $M_\phi$ are subject to theoretical and experimental constraints. For $\lambda_3$, it is obvious in the left panel of Fig.~\ref{figphi} ($\lambda_4=0$) that in some region of the parameter space $\lambda_3$ can not be too large, or $\lambda$ becomes non-perturbative at energies much lower than the Planck scale. If we consider further the vacuum stability, $\lambda_3$ is more strictly constrained\footnote{If we require the potential is bounded from blow, $\lambda_3$ has to satisfy the relation $\lambda_3^2 < 2\lambda\lambda_1$, when the scalar $\kappa$ is turned off. Furthermore, even we have $\lambda_1 >0$ at the electroweak scale, according to the RGE running, $\lambda_1$ may become negative at higher energy scales, destabilizing the vacuum. It is a bit non-trivial to stabilize the vacuum in the Zee-Babu model.}.
For simplicity, we na\"ively assume $|\lambda_3| < 4\pi$.

In high energy collider experiments, the scalar $\phi$ in the Zee-Babu model behaves like a charged Higgs with weak isospin $I_3=0$ whose Yukawa interaction is solely the coupling to left-handed leptons. The experiments ALEPH, DELPHI and L3 at LEP have searched for charged Higgs~\cite{phi-at-lep1}. They considered the two or three decay channels $H^+ \rightarrow \tau^+\nu_\tau,\,c\bar{s},\,W^\ast A$ ($A$ being light pseudoscalar), giving a limit of 79.3~GeV at 95\% confidence level, which is independent of the branching ratio ${\rm BR}(H^+ \rightarrow \tau^+\nu_\tau)$. By analyzing the dilepton events with missing transverse momentum, the experiment OPAL obtained a more stringent limit of 92~GeV at 95\% confidence level~\cite{phi-at-lep2}, assuming the charged Higgs decaying totally to tau leptons plus missing energy. If the Zee-Babu model is assumed to produce the tiny neutrino masses, either for normal hierarchy (NH) or inverted hierarchy (IH) of neutrino spectrum, the branching ratios of $\phi$ are determined by the three neutrino mixing angles $\theta_{ij}$ and the Dirac CP violating phase $\delta$~\cite{zb-ph1,zb-ph2},
\begin{eqnarray}
{\rm NH:} \quad &&
\frac{f_{e\tau}}{f_{\mu\tau}} =
\frac{\tan\theta_{12}\cos\theta_{23}}{\cos\theta_{13}}+\tan\theta_{13}\sin\theta_{23}e^{-i\delta}\ ,\nonumber \\
&& \frac{f_{e\mu}}{f_{\mu\tau}} = \frac{\tan\theta_{12}\sin\theta_{23}}{\cos\theta_{13}}-\tan\theta_{13}\cos\theta_{23}e^{-i\delta}\ ,\\
{\rm IH:} \quad &&
\frac{f_{e\tau}}{f_{\mu\tau}} =
-\frac{\sin\theta_{23}}{\tan\theta_{13}}e^{-i\delta}\ ,\nonumber \\
&&\frac{f_{e\mu}}{f_{\mu\tau}} =
\frac{\cos\theta_{23}}{\tan\theta_{13}}e^{-i\delta}\ .
\end{eqnarray}
If we utilize the OPAL result to constrain $M_\phi$, the limit has to be brought a bit down. Nevertheless, as the neutrino spectrum is not yet pinned down, we conservatively take the orginal OPAL limit of 92~GeV as the lower bound for $M_\phi$. Charged Higgs has also been searched for at hadron colliders with higher energies in decays of top quark $t \rightarrow bH^+$~\cite{phi-at-hadronic-colliders}. However, since the extra scalars $\phi, \kappa$ do not couple to quarks, these searches are inapplicable for the Zee-Babu model. Therefore, the collider constraint on $\phi$ mass is taken as $M_\phi >92$~GeV, at 95\% confidence level.

We show in the left panel of Fig.~\ref{figphi} the plots of the enhancement ratios $R_{\gamma\gamma}^\phi=1.1,\,1.3,\,1.5$, $\,1.7,\,2$ for $h \rightarrow \gamma\gamma$ rate, as functions of $\lambda_3$ and $M_\phi$. The blue region on the left is the excluded region $M_\phi > 92$~GeV. To interfere constructively with the SM contribution, the value of $\lambda_3$ is negative, as aforementioned. It is clear from the figure that the large ratio $R_{\gamma\gamma}$ can be easily achieved for a light scalar $\phi$ with mass of hundreds of GeV and a negative coupling $\lambda_3 \sim -3$.

If $\lambda_3 > 0$, the contribution of the scalar $\phi$ to $h \rightarrow \gamma\gamma$ decay rate interferes destructively with that from the SM $W$ boson. To obtain the large $h \rightarrow \gamma\gamma$ width enhancement observed at LHC, $M_\phi$ has to be close to but larger than the collider constraint, i.e. 92~GeV, and meanwhile the magnitude of $\lambda_3$ is large $\sim$10, i.e. close to its perturbativity bound. Therefore the case with positive $\lambda_3$ is not phenomenologically favored.

\subsection{$\phi$ off, $\kappa$ on}

In this subsection, we consider the case in which $\phi$ is turned off (or this can be understood as $\lambda_3=0$ at the electroweak scale), and $h \rightarrow \gamma\gamma$ receives only the contribution from the scalar $\kappa$ besides the SM particles. The scalar $\kappa$ carries double electric charges, and the corresponding enhancement factor is, analogous to Eq.~(\ref{phi}),
\begin{eqnarray}
R_{\gamma\gamma}^\kappa = \left| 1+  \frac{1}2 \lambda_4 \frac{v^2}{M_\kappa^2}\frac{4A_0(\tau_{\kappa})}{ A_1(\tau_W)+ N_c Q_t^2 \, A_{1/2}(\tau_t)}\right|^2\,.
\end{eqnarray}
The constraint on $\lambda_4$ from perturbativity is similar to that for $\lambda_3$ discussed in the previous subsection, i.e. $|\lambda_4| < 4\pi$, regardless of the difference of electric charges carried by $\phi$ and $\kappa$.

At colliders, the scalar $\kappa$ acts like ``right-handed'' doubly charged Higgs with weak isospin $I_3=0$ (``right-handed'' means that the scalar couples to right-handed charged leptons or left-handed anti- charged leptons). Doubly charged Higgs has been searched for at LEP~\cite{kappa-at-lep} and hadron colliders~\cite{kappa-at-tevatron,kappa-at-CMS}, through the process $q\bar{q} \rightarrow \gamma^\ast/Z^\ast \rightarrow H^{++}H^{--} \rightarrow \ell^+_a\ell^+_b\ell^-_c\ell^-_d$ ($a,\,b,\,c,\,d = e,\,\mu,\,\tau$), including lepton flavor violating decays, and the strongest bound is from CMS at the LHC~\cite{kappa-at-CMS}. In the inclusive search by CMS, they considered all the six decay channels, i.e. ${\rm BR}(H^{++} \rightarrow e^+e^+/ \mu^+\mu^+/ \tau^+\tau^+/ e^+\mu^+/ e^+\tau^+/ \mu^+\tau^+) =100\%$, as well as four benchmark points with different branching ratios, such as
\begin{eqnarray}
&&{\rm BR}(H^{++} \rightarrow e^+e^+) =1/3\,, \nonumber\\
&&{\rm BR}(H^{++} \rightarrow \mu^+\mu^+) =1/3\,,\nonumber\\
&&{\rm BR}(H^{++} \rightarrow \tau^+\tau^+) =1/3\,.
\end{eqnarray}
Based on the pair production analysis only, the exclusion limit on the mass of the doubly charged Higgs varies from 165~GeV to 391~GeV, largely depending on the decay mode. In the Zee-Babu model, the Yukawa couplings of the doubly charged scalar $\kappa$ to charged leptons, $g_{\alpha\beta}$, are related to the masses of charged leptons and neutrino oscillation parameters, and are constrained by low energy lepton flavor violating processes, e.g. $\tau^- \rightarrow \mu^+\mu^-\mu^-$~\cite{zb-ph1,zb-ph2}. The values of the couplings $g_{\alpha\beta}$ are yet undetermined, we choose the weakest limit on doubly charged Higgs from CMS as the mass limit for $\kappa$: $m_{\kappa} > 165$~GeV.

It is worth to mention that the doubly charged Higgs in the CMS analysis is assumed to be from the type-\uppercase\expandafter{\romannumeral2} seesaw model~\cite{seesawII} and consequently has weak isospin $I_3 = 1$, which is different from the ``right-handed'' scalar $\kappa$ with $I_3=0$. Due to their different coupling to the virtual $Z$ boson, the pair production cross section of right-handed doubly charged scalars is somewhat smaller than that of the left-handed ones~\cite{kappa-at-tevatron,kappa-left-right}. When we utilize the mass limit originally for the left-handed doubly charged scalar to constrain the mass of right-handed ones, we should properly scale down the pair production cross section, thus reduce the mass limits. However, to compromise to some extent the dependence of the mass limit on the branching ratios of the doubly charged Higgs, we still use the original value of 165~GeV from CMS as our limit for $m_\kappa$.

The plots of the enhancement ratios $R_{\gamma\gamma}^\kappa=1.1,\,1.3,\,1.5,\,1.7,\,2$ as functions of $\lambda_4$ and $m_\kappa$ are shown in the right panel of Fig.~\ref{figphi}. The blue region on the left is the constraint $m_\kappa > 165$~GeV from CMS. As the electric charge carried by the scalar is doubled, the allowed region for the scalar mass and the quartic coupling is somewhat larger, as we expect, compared to that for $\phi$.

If $\lambda_4$ is positive, due to the stringent bound on the scalar mass, as in the case for $\phi$, to obtain the large enhancement ratio $R_{\gamma\gamma}$, the scalar mass $m_\kappa$ has to be close to but larger than its experimental bound, and $\lambda_4$ is very large $\lambda_4 \simeq 10$, which is disfavored phenomenologically.

\subsection{$\phi$ on, $\kappa$ on}

Now we turn on both scalar fields $\phi$ and $\kappa$. In this case, the enhancement ratio is
\begin{eqnarray}
R_{\gamma\gamma} = \left| 1+
\frac{1}2 \lambda_3 \frac{v^2}{M_\phi^2}\frac{A_0(\tau_{\phi})}{ A_1(\tau_W)+ N_c Q_t^2 \, A_{1/2}(\tau_t)}
+\frac{1}2 \lambda_4 \frac{v^2}{M_\kappa^2}\frac{4A_0(\tau_{\kappa})}{ A_1(\tau_W)+ N_c Q_t^2 \, A_{1/2}(\tau_t)}\right|^2\,.
\end{eqnarray}
Collider constraints on $m_\phi$ and $m_\kappa$ and the bound on $\lambda_3$ and $\lambda_4$ from perturbativity remain the same.

\begin{figure}[t]
\centering
\includegraphics[width=12cm]{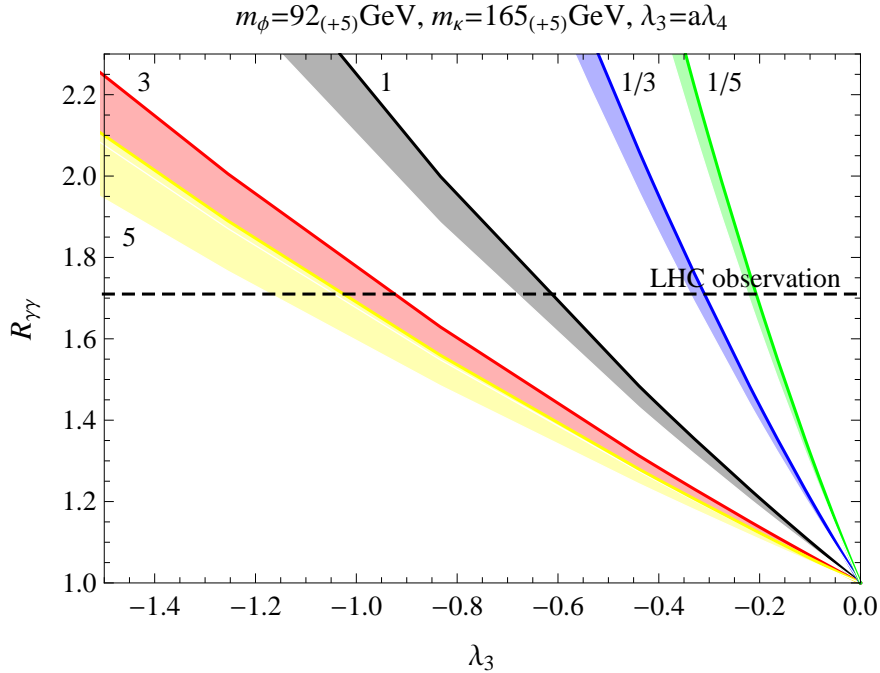}
\caption{Enhancement ratios $R_{\gamma\gamma}$ as function of $\lambda_3$ are shown as the yellow, red, black, blue and green curves, when we set $m_{\phi} = 92_{(+5)}$~GeV, $m_{\kappa} = 165_{(+5)}$~GeV, and parametrize $\lambda_3 = a\lambda_4$, with $a=5,~3,~1,~1/3,~1/5$ in the plot. The lighter bands are due to the variations of the scalar masses. The dashed black horizontal line denotes the large enhancement ratio $R_{\gamma\gamma}$ recently observed at the LHC.}
  \label{figr}
\end{figure}

We first consider the phenomenologically most favored case in which the two couplings $\lambda_{3,\,4}$ are both negative, i.e. $-4\pi< \lambda_{3,\,4}<0$, to have contributions interfering constructively with the SM $W$ boson. With collider constraints on the scalar masses, $m_{\phi} > 92$~GeV and $m_{\kappa} > 165$~GeV, we now show the plots for the parameter regions allowed to obtain the large enhancement ratio $R_{\gamma\gamma}$.
We first assume the scalar masses to be close to their experimental bounds. More specifically, $m_\phi$ and $m_\kappa$ are, respectively, less than 5~GeV larger than their collider lower bounds, denoted as $m_{\phi} = 92_{(+5)}$~GeV and $m_{\kappa} = 165_{(+5)}$~GeV. We parametrize the two quartic couplings as $\lambda_3 = a\lambda_4$, where $a=5,\,3,\,1,\,1/3,\,1/5$. The enhancement ratios $R_{\gamma\gamma}$ as functions of $\lambda_3$ is shown in Fig.~\ref{figr} as the yellow, red, black, blue and green curves. The lighter bands are due to the variations of the scalar masses in the regions $92<m_\phi<97$~GeV and $165<m_\kappa<170$~GeV. The horizontal dashed black line is the enhancement ratio recently observed at the LHC. Obviously, as the two scalars are relatively light, $|\lambda_{3,\,4}|$ are both very small, $\sim \mathcal{O}(0.1-1)$.

\begin{figure}[htbp]
  \includegraphics[width=6.6cm]{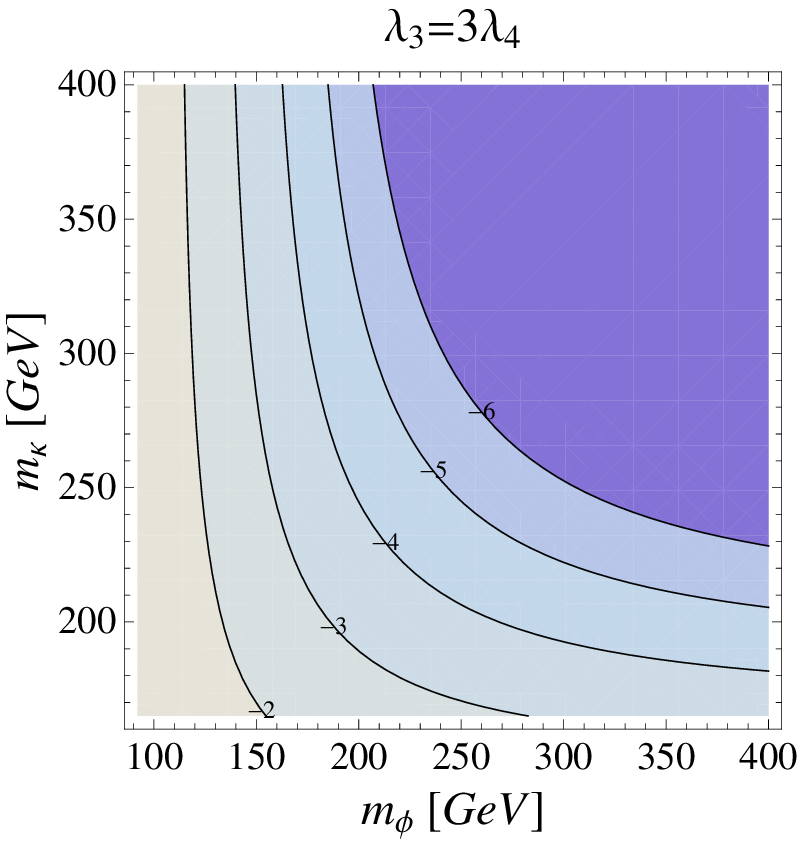}  \hspace{.5cm}
  \includegraphics[width=6.6cm]{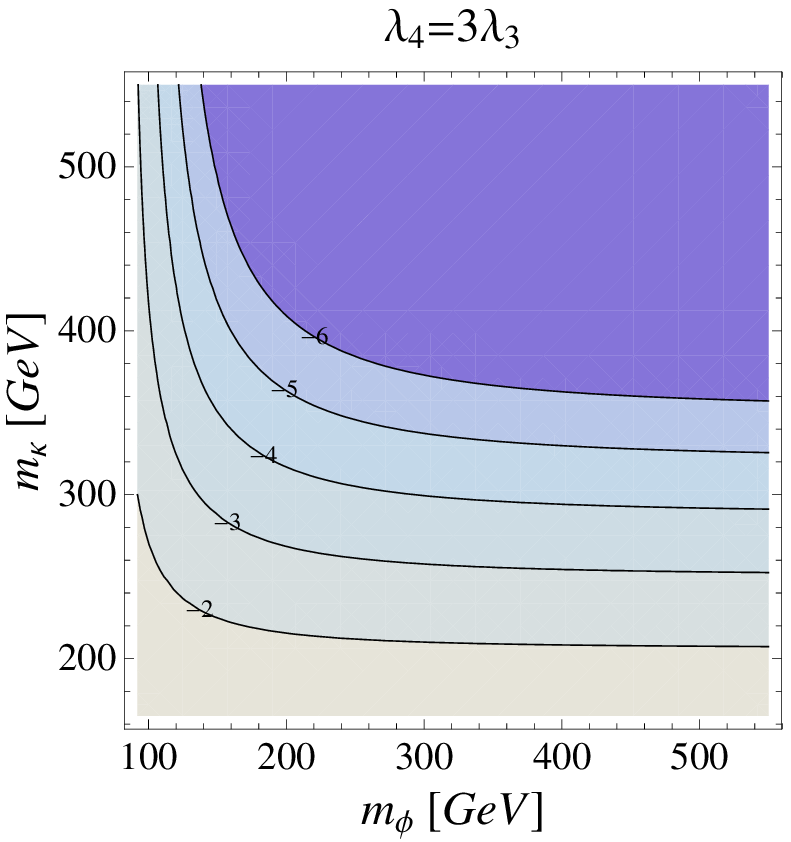}
  \caption{Left panel: Contours of $\lambda_3$ required to obtain the enhancement ratio $R_{\gamma\gamma} =1.71$ as functions of $m_\phi$ and $m_\kappa$, with $\lambda_3 = 3\lambda_4$;
  right panel: Contours of $\lambda_4$ with $\lambda_4 = 3\lambda_3$.}
  \label{figm1}
\end{figure}

When $\phi$ and $\kappa$ are heavier, i.e. with masses of hundreds of GeV, $|\lambda_{3,\,4}|$ should be somewhat large to produce a large $R_{\gamma\gamma}$, as seen in Fig.~\ref{figm1}, which shows in the left (right) panel the contour of $\lambda_3~ (\lambda_4)$ required to obtain the enhancement ratio $R_{\gamma\gamma} =1.71$ as a function of $m_\phi$ and $m_\kappa$, with couplings $\lambda_3 = 3\lambda_4 ~(\lambda_4 = 3\lambda_3)$. When both $\lambda_{3,\,4}$ are positive, the situation is a bit better than the case with solely one of the scalars, but to produce a large enhancement ratio $R_{\gamma\gamma}$, $\phi$, $\kappa$ are both relatively light and $\lambda_{3,\,4}$ are both large, $\lambda_{3,\,4} \gtrsim 5$. This is still not the phenomenologically favored case.

We have to mention that it is almost impossible to find the parameter space where all the Higgs vacuum stability, perturbativity and enhanced Higgs diphoton decay rate can be achieved simultaneously. To fulfill the Higgs vacuum stability and perturbativity constraints, scalar couplings should be small. To achieve the demanded $h\to \gamma \gamma$ enhancement factor, $\lambda_{3,4}$ should be relatively large. This paradox seems improbable to be resolved without introducing new ingredients into the model. 

\section {Searches of $\phi$, $\kappa$ at the LHC}

\begin{figure}[htbp]
\vspace*{4.5cm}
\hspace*{1.5em}
\includegraphics[width=0.4\textwidth]{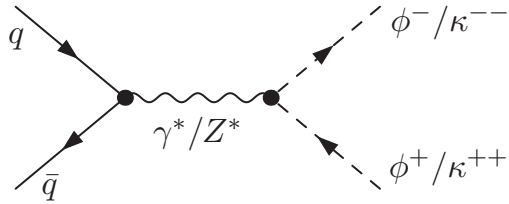}
\caption{$\phi/\kappa$ pair production via $q\bar q$ annihilation.}
\vspace{4cm}
\label{qqbardiagram}
\end{figure}

\begin{figure}[t]
\hspace*{-4em}
\includegraphics[width=0.51\textwidth]{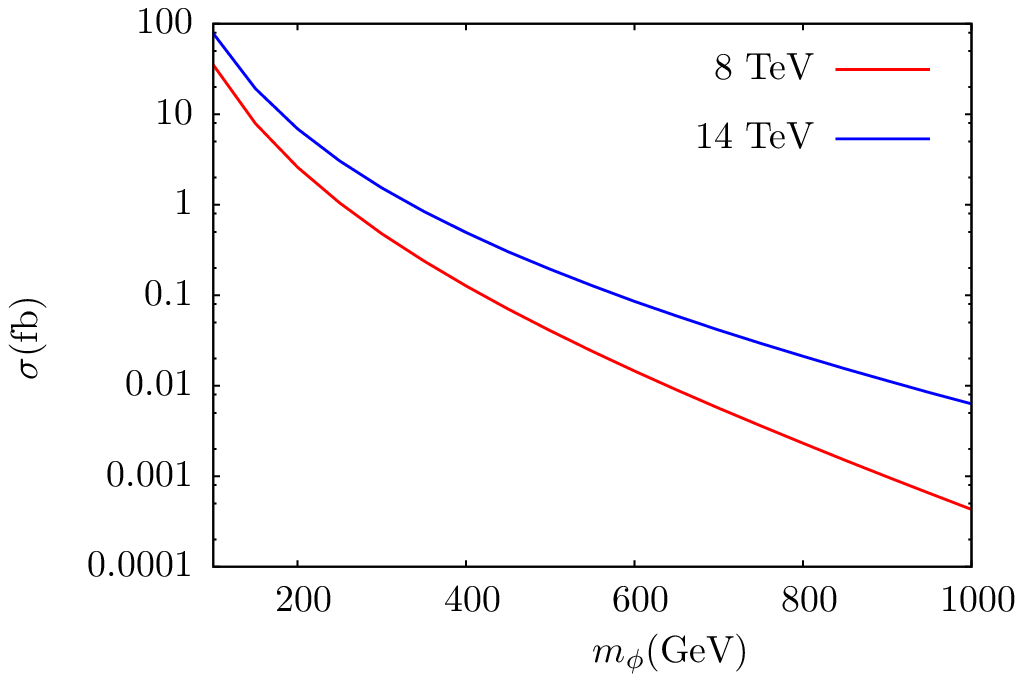}
\hspace{-1em}
\includegraphics[width=0.5\textwidth]{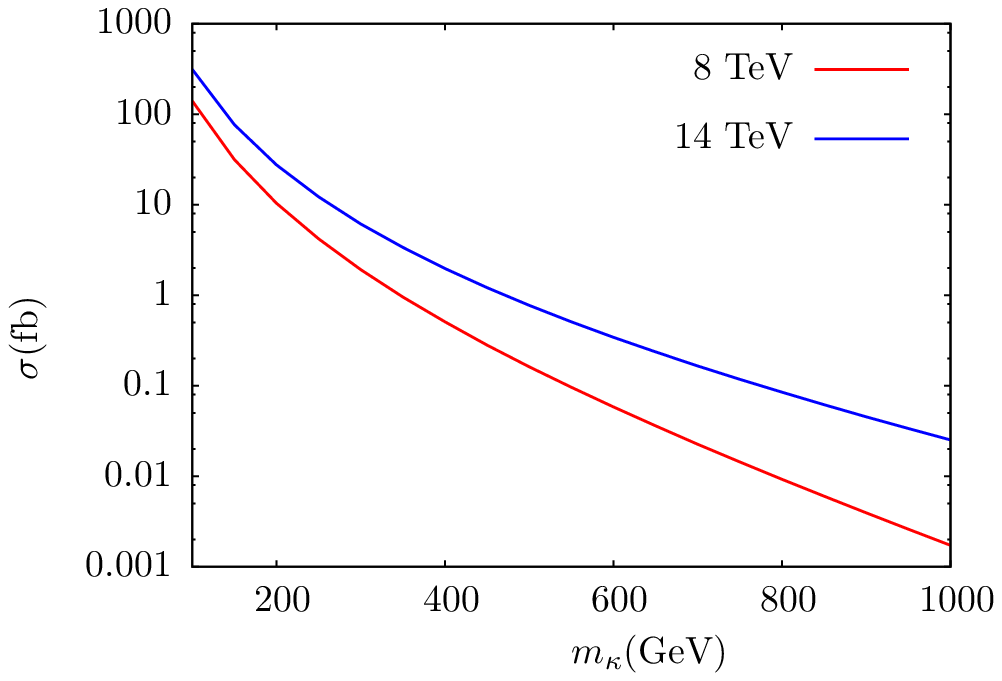}
\caption{$\phi/\kappa$ pair production cross section at the LHC. }
\label{xsecphikappa}
\end{figure}

In this section we briefly sketch the search for $\phi$, $\kappa$ at the LHC. The phenomenology of extra scalars in the Zee-Babu model at the LHC has been analyzed in Ref.~\cite{zb-ph2}. Most of the discussions in that reference essentially apply here. Recall that the extra scalar fields $\phi$ and $\kappa$ do not couple to quarks, the main production of $\phi$ and $\kappa$ pairs at the LHC is via quark-antiquark annihilation (see Fig.~\ref{qqbardiagram}). An important feature of this production mechanism is that the cross section depends only on one unkonwn parameter, the mass of $\phi$ or $\kappa$. The mass of $\phi$ and $\kappa$ has been constrained to be roughly above 100 GeV by LEP searches for charged scalars. In Fig.~\ref{xsecphikappa} we plot the LO $\phi/\kappa$ pair production cross section at 8 and 14 TeV at the LHC, as a function of their mass. As shown in the figure, the cross section decreases quite rapidly with the scalar mass $m_{\phi,\, \kappa}$, from a few hundred fb for $m_{\phi,\, \kappa} =100$\,GeV to less than $0.1$\,fb for $m_{\phi,\, \kappa} =1$\,TeV. The pair production cross section for $\kappa$ is larger than that for $\phi$ with the same mass, because $\kappa$ is doubly charged.

After being pair produced, $\phi$ and $\kappa$ will decay to lighter particles~\footnote{We assume these extra scalars are not long-lived, indeed the low energy phenomenology and neutrino data imply that $\kappa$ cannot be a long-lived particle~\cite{zb-ph2}.}. $\phi$ decays to charged lepton and neutrino, whereas $\kappa$ can decay to a pair of $\phi$ provided that $m_\kappa>2m_\phi$, or to a pair of charged leptons. Due to the presence of neutrino in the decay product of $\phi$, and also due to its smaller production cross section, the discovery potential for $\phi$ is less than that for $\kappa$. A clean and distinct signature for $\kappa$ is provided by its decay to electron/muon pair with high invariant mass, where the detection of both electrons and muons is quite efficient. To estimate the number of events in these decay channels, we take the optimistic scenario in Ref.~\cite{zb-ph2} where all produced $\kappa$ are assumed to decay to electron/muon pair. In this case, we have 207 (3) events for an estimated integrated luminosity 20\,$fb^{-1}$ at $\sqrt s=8$ TeV with $m_\kappa=200$\,GeV ($m_\kappa=500$\,GeV), and 8264 (231) events for an integrated luminosity 300\,$fb^{-1}$ at $\sqrt s=14$ TeV with $m_\kappa=200$\,GeV ($m_\kappa=500$\,GeV).

\section{conclusion}

If the observed $\sim125~{\rm GeV}$ boson is, indeed, the Higgs boson, then stability of the SM electroweak vacuum up to the Planck scale may require that some BSM degrees of freedom become active above the TeV scale.  The enhanced Higgs to diphoton decay rate also points to new physics beyond the SM. In this paper we studied the Higgs vacuum stability and enhanced Higgs to diphoton decay rate in the Zee-Babu model, which was used to generate tiny Majorana neutrino masses at two-loop level. We find that it is rather difficult to find overlapping regions allowed by the vacuum stability and diphoton enhancement constraints. As a consequence, it is almost inevitable to introduce new ingredients into the model, in order to resolve these two issues simultaneously. We also calculated the production cross section of new charged scalars at the CERN LHC, which does not change much even if new degrees of freedom are introduced into the model. If the model could be made compatible with current Higgs diphoton decay rate observation by e.g. introducing new degrees of freedom, it would be possible to detect these scalars at the LHC in the near future.

\begin{acknowledgments}
This work was supported in part by DOE contracts DE-FG02-08ER41531 and the Wisconsin Alumni Research Foundation (WCHAO).

\end{acknowledgments}

\end{document}